\documentstyle[12pt,a4wide]{article}
\input epsf
\newcommand{\be}{\begin{equation}}
\newcommand{\ee}{\end{equation}}
\newcommand{\pr}{\partial}
\newcommand{\I}{{\cal I}}
\newcommand{\bpi}{\mbox{\boldmath $\pi$}}
\newcommand{\pauli}{\mbox{\boldmath $\tau$}}
\font\mybb=msbm10 at 11pt

\def\bb#1{\hbox{\mybb#1}}

\def\bR {\bb{R}}

\newcommand{\news}{\setcounter{equation}{0}}
\def\ben{\begin{equation}}
\def\een{\end{equation}}
\def\bea{\begin{eqnarray}}
\def\eea{\end{eqnarray}}
\begin{document}

\title{\vskip -0pt
\bf \large \bf SKYRMIONS AND THE PION MASS\\[30pt]
\author{Richard A. Battye$^{1}$
and Paul M. Sutcliffe$^{2}$\\[10pt]
\\{\normalsize $^{1}$
{\sl Jodrell Bank Observatory, Macclesfield, Cheshire SK11 9DL U.K.}}
\\{\normalsize {\sl $\&$  School of Physics and Astronomy,
Schuster Laboratory,}}
\\{\normalsize {\sl University of Manchester, Brunswick St,
 Manchester M13 9PL, U.K.}}
\\{\normalsize {\sl Email : rbattye@jb.man.ac.uk}}\\
\\{\normalsize $^{2}$  {\sl Institute of Mathematics,
University of Kent, Canterbury, CT2 7NF, U.K.}}\\
{\normalsize{\sl Email : P.M.Sutcliffe@kent.ac.uk}}\\}}
\date{October 2004}
\maketitle
 
\begin{abstract}
We present numerical evidence that suggests  the introduction of a
non-zero pion mass might dramatically affect  the structure of minimal
energy Skyrmions.  It appears that the shell-like Skyrmions which are
the minima when the pions are massless can fail to be minimal energy
bound states for particular baryon numbers, with a strong dependence
upon the value of the pion mass.  The effects of a pion mass may include
the replacement of shell-like configurations with crystal chunks   and
the loss of shell-like bound states with baryon numbers five
and eight; which is in agreement with expectations based on  real nuclei.
\end{abstract}

\newpage
 
\section{Introduction}\news
The Skyrme model \cite{Sk} is a nonlinear theory of pions which is
an approximate, low energy effective theory of quantum chromodynamics,
obtained in the limit of a large number of quark colours \cite{Wi}.
Skyrmions are topological soliton solutions of the model and 
are candidates for an effective description of nuclei, with an identification
between soliton and baryon numbers. 

The Lagrangian of the Skyrme model contains only three free parameters;
two of these set the energy and length units and the third corresponds 
to the (tree-level) pion mass. In Ref.\cite{ANW} the energy and length
units were calculated by fitting to the masses of the proton and delta
resonance assuming massless pions, and in Ref.\cite{AN} this calculation
was repeated using the measured value for the pion mass. As the pion mass
is relatively small its inclusion modifies the computed energy and length
units by only a few percent in this approach.
The main effect of the pion mass is that the Skyrmion becomes exponentially
localized, rather than the algebraic asymptotic behaviour of the field 
in the massless pion model. 

It is often said that the pion mass has
little qualitative effect and so it is frequently neglected in the
study of Skyrmions. For more than two Skyrmions there are very few 
studies that include massive pions \cite{BTC,BBT,Kop} and furthermore 
these results merely reinforce the view that there are few 
qualitative differences, since the Skyrmions show no substantial variations
from those in the massless pion model \cite{BS3}. 
In addition, as far as we are aware, there are no studies of 
the properties of multi-Skyrmions as a function of the pion mass.
In this paper
we show that the inclusion of a pion mass may have dramatic effects,
if either the baryon number or the pion mass is not small. 

In section \ref{sec-standard} we discuss the properties of  shell-like 
Skyrmions with baryon numbers
upto 40, using the standard value of the pion mass, as calculated
in Ref.\cite{AN}. We show that this pion mass leads to an 
interesting modification to the asymptotic behaviour of the energy 
per baryon at large baryon number.
Above baryon number 12 this quantity starts to increase, 
in constrast to the decreasing trend in the massless case.
This means that for large enough baryon numbers there can be no
shell-like bound states; for example, the charge 24 Skyrmion shell
is unstable to the decay into two charge 12 Skyrmions. 
This is because the pion mass produces 
an energy contribution which depends upon the
volume enclosed within the shell of the Skyrmion.

Candidate replacements for shell-like Skyrmions involve chunks of the
Skyrme crystal.
 The most natural crystal chunk occurs
at charge 32, so in section \ref{sec-32} we compare the energy of this
crystal chunk with the minimal energy charge 32 shell, as a function of the
pion mass. For massless pions the shell has lowest energy, but at a value
well below the standard pion mass we find that there is a cross-over so
that the crystal chunk has lowest energy.

When the pion mass is included it is invariably set to the 
value calculated in Ref.\cite{AN}, which we are referring to as 
the standard value. As we have already pointed out, 
this calculation involves first determining
the energy and length units by fitting to the masses of the proton
and delta resonance, and then converting the measured pion mass into these
units. At the time of this calculation there were no results available
for multi-Skyrmions, so this approach was essentially the only one
available. However, there are now substantial results available (at
least classically) for a large number of multi-Skyrmions, and it is
our opinion that it is worth reconsidering whether fitting to an unstable
resonance is the best way to determine the parameters of the Skyrme model,
when the main application is to try and model nuclei of all baryon numbers. 

With this point of view in mind there is clearly 
more flexibility for the physical
value of the pion mass, since we are prepared to sacrifice the approximation
of the delta resonance in favour of a better fit to nuclei with
baryon number greater than one. This motivates our study in section 
\ref{sec-large}, where we consider shell-like Skyrmions with baryon
numbers one to eight, and pion masses from zero upto a value several times
larger than the standard one. We find evidence for some interesting 
behaviour at relatively large pion masses. Notably, there appears to
be a range of pion masses at which the shell-like Skyrmions with baryon numbers
 five and eight are no longer bound states.  
This result is in agreement with real nuclei, since there are no stable 
states with five or eight nucleons.

Finally, in section \ref{sec-conclusion} we summarize our findings and
describe the future work that needs to be done to investigate the
validity of some of our conjectures, and to test our ideas regarding 
the comparisons with real nuclei.

\section{Skyrmions}\news\label{sec-skyrmions}
\subsection{The Skyrme model}
The field $U$ of the Skyrme model \cite{Sk} 
is an $SU(2)$-valued scalar with the 
Lagrangian  
\be
L = \int\Biggl\{\frac{F_\pi^2}{16}\mbox{Tr}(\pr_\mu U\pr^\mu U^\dagger)
+\frac{1}{32e^2}\mbox{Tr}([\pr_\mu U U^\dagger,\pr_\nu U U^\dagger]
[\pr^\mu U U^\dagger,\pr^\nu U U^\dagger])
+\frac{m_\pi^2F_\pi^2}{8}\mbox{Tr}(U-1)\Biggr\}\, d^3x.
\label{skylagconstants}
\ee
Here $F_\pi$, $e$ and $m_\pi$ are parameters, whose values are
fixed by comparison with experimental data.
$F_\pi$ may be interpretated as the pion decay constant, and so
could be fixed by its experimental value, but this 
approach is not usually taken (see for example Refs.\cite{ANW,AN}).
$e$ is a dimensionless constant and $m_\pi$ is the pion mass,
which if set to be non-zero is generally taken to be the 
measured value once $F_\pi$ and $e$ are fixed in some way.

The parameters $F_\pi$ and  $e$ can be scaled away by using energy and 
length units of ${F_\pi}/{4e}$ and ${2}/{eF_\pi}$ respectively, which we adopt
from now on. In terms of these units the Skyrme Lagrangian
becomes
\be
L=\int \left\{-\frac{1}{2}\mbox{Tr}(R_\mu R^\mu)+\frac{1}{16}
\mbox{Tr}([R_\mu,R_\nu][R^\mu,R^\mu])+m^2\mbox{Tr}(U-1)
\right\} \, d^3x,
\label{skylag}
\ee
where we have introduced the $su(2)$-valued current
 $R_\mu=(\partial_\mu U)U^\dagger,$ and the normalized 
pion mass $m=2m_\pi/(F_\pi e).$

From Ref.\cite{AN} the standard values used are $F_\pi=108\mbox{MeV}$,
 $e=4.84$ and $m_\pi=138\mbox{MeV},$ giving $m=m_*=0.526.$ 
So, if the energy and length units are fixed by fitting the masses
of the proton and delta resonance then the pion mass in our 
units is $m_*.$ 
However, if a different approach is taken to determining
$F_\pi$ and $e,$ for example, by requiring a fit to the properties
of nuclei with baryon numbers greater than one, then a pion mass
other than $m_*$ would result. This motivates our later
study in which we will vary $m.$ 

The boundary condition is that $U\rightarrow 1$ at spatial infinity,
and such maps are characterized by an integer-valued winding number.
This topological charge is interpreted as the baryon number 
and has the expression
\be
B=-{1\over 24\pi^2}\int \epsilon_{ijk}
 {\rm Tr}\left(R_iR_jR_k\right)\, d^3x.
\label{baryon}
\ee
In this paper we are only concerned with static fields, in which 
case the Skyrme energy derived from the Lagrangian (\ref{skylag}) is
\be
E=\frac{1}{12\pi^2}\int \left\{-{1 \over 2}\mbox{Tr}(R_iR_i)-{1 \over 16}
\mbox{Tr}([R_i,R_j][R_i,R_j])+m^2\mbox{Tr}(1-U)\right\} \, d^3x\,,
\label{skyenergy}
\ee
where we have introduced the additional factor of $1/12\pi^2$ for later
 convenience. With this normalization of the energy 
the Faddeev-Bogomolny bound reads $E\ge |B|.$

To make contact with the nonlinear pion theory $U$ is written as
\be
U=\sigma +i\bpi\cdot\pauli
\ee
where $\pauli$ denotes the triplet of Pauli matrices,
$\bpi=(\pi_1,\pi_2,\pi_3)$ is the triplet of pion fields and $\sigma$
is determined by the constraint $\sigma^2+\bpi\cdot\bpi=1.$

\subsection{The rational map ansatz and energy prediction}
For massless pions the minimal energy Skyrmions have been obtained
for all $B\le 22$ \cite{BS3}. For low charges the Skyrmions are
typically very symmetric (often having Platonic symmetries) and agree 
qualitatively with the Skyrmions in the case of the standard pion mass
\cite{BTC}, though only charges upto five have been computed accurately
in this latter case. For higher charges the Skyrmions become even more 
shell-like, with the baryon and energy densities being localized on the
edges of polyhedra which are generically of the fullerene type.
All these known minimal energy Skyrmions (and some others) 
can be very well approximated using the rational map ansatz \cite{HMS},
which we now briefly review.  

In this approach a Skyrme field with baryon number $B$ is 
constructed from a degree
$B$ rational map between Riemann spheres. Although this
ansatz does not give exact solutions of the static Skyrme equations,
it produces approximations which have energies only a few percent
above the numerically computed solutions.
Briefly, use spherical coordinates in $\bR^3$, so
that a point ${\bf x}\in\bR^3$ is given  by a pair $(r,z)$, where
$r=\vert{\bf x}\vert$ is the  distance from the origin, and $z$ is a
Riemann sphere coordinate giving the point on the unit two-sphere
which intersects the half-line through the origin and the point ${\bf
x}$.
Now, let $R(z)$ be a degree $B$ rational map between Riemann spheres,
that is, $R=p/q$ where $p$ and $q$ are polynomials in $z$ such that
$\max[\mbox{deg}(p),\mbox{deg}(q)]=B$,  and $p$ and $q$ have no common
factors.  Given such a rational map the ansatz for the Skyrme field is
\be  U(r,z)=\exp\bigg[\frac{if(r)}{1+\vert R\vert^2} \pmatrix{1-\vert
R\vert^2& 2\bar R\cr 2R & \vert R\vert^2-1\cr}\bigg]\,,
\label{rma}
\ee where $f(r)$ is a real profile function satisfying the  boundary
conditions $f(0)=\pi$ and $f(\infty)=0$, which is determined by
minimization of the Skyrme energy of the field (\ref{rma}) given a
particular rational map $R$. 

Substitution of the rational map ansatz (\ref{rma}) into the Skyrme
energy (\ref{skyenergy}) 
results in the following expression 
 \be
E=\frac{1}{3\pi}\int \bigg( r^2f'^2+2B(f'^2+1)\sin^2 f+\I\frac{\sin^4
f}{r^2}+2m^2r^2(1-\cos f)\bigg) \ dr\,,
\label{rmaenergy}
\ee 
where $\I$ denotes the integral 
\be \I=\frac{1}{4\pi}\int \bigg(
\frac{1+\vert z\vert^2}{1+\vert R\vert^2}
\bigg\vert\frac{dR}{dz}\bigg\vert\bigg)^4 \frac{2i \  dz  d\bar z
}{(1+\vert z\vert^2)^2}\,.
\label{i}
\ee 
To minimize the energy (\ref{rmaenergy}) one first
determines the rational map which minimizes $\I$, 
then given the minimum value of $\I$ it is a simple exercise to find the
minimizing profile function.
Thus, within the rational map ansatz, the problem of finding
the minimal energy Skyrmion reduces to the simpler problem of
calculating the rational map which minimizes the function
$\I$. Note that since $\I$ is independent of $m$ then the same
minimizing maps for massless pions will also be the minimizing maps
for all values of the pion mass. These maps have been determined 
numerically \cite{BHS} for all $B\le 40.$ The effect of the pion mass
is merely to change the profile function $f(r)$ and hence the energy.

Typically the energy of the rational map ansatz overestimates the
true energy of the Skyrmion by a few percent (except for $B=1$ where
it is exact). A simple, but as it turns out quite accurate, method
to predict the energy of a given Skyrmion as a function of the pion mass
is to assume that the percentage error is independent of the pion mass.
Let $E(m)$ denote the energy of the rational map ansatz for a given
Skyrmion with pion mass $m.$ Then if the true energy at
zero pion mass, denoted by $E_0,$ is known 
we define the predicted energy to be
\be
\widetilde E(m)=E(m)\frac{E_0}{E(0)}.
\label{prediction}
\ee
As we shall show in later sections, by comparison with the
results of full field simulations, this simple prediction method
gives a surprisingly accurate approximation to the energy of the given
Skyrmion with pion mass $m.$

\subsection{The Skyrme crystal}
Studies of an hexagonal Skyrme lattice \cite{BS4} suggest that 
for massless pions the
limiting energy per baryon for large shell-like fullerene Skyrmions
is $1.061,$ and this is consistent with results obtained for 
icosahedral Skyrmions with large baryon numbers \cite{BHS}. 
However, this is larger than the value $1.036$ obtained for
the infinite triply periodic Skyrme crystal \cite{KSh,CJJVJ,BS4}, again
for massless pions, which has the lowest energy per baryon
of any known Skyrme field.

The Skyrme crystal has cubic symmetry and is composed 
of half-Skyrmions. A good approximation, which has the 
correct symmetry structure,
is given by the formulae \cite{CJJVJ}
\be
\sigma=s_1s_2s_3\,, \quad
\pi_1=c_1\sqrt{1-\frac{c_2^2}{2}-\frac{c_3^2}{2}
+\frac{c_2^2c_3^2}{3}} \hskip 1cm \mbox{and cyclic} \,,
\label{crystal}
\ee
where $s_i=\sin(\pi x_i/L)$ and $c_i=\cos(\pi x_i/L)$.
Here $2L$ is the side length of the cube forming the unit
cell of the crystal, and the energy is minimal when $L=4.7.$

The Skyrme crystal, and the approximation (\ref{crystal}),
 will play a role in section \ref{sec-32}.

\section{Shells with a pion mass}\news\label{sec-standard}

\begin{table}
\centering
\begin{tabular}{|c|c|c|c|}
\hline
$B$ &  $E_0/B$ & $E_{m_*}/B$ & $\widetilde E(m_*)/B$\\\hline   
1 &	1.2322&	1.3119&  1.3077\\
4&	1.1200&	1.1926&  1.1888\\
8&	1.0960&	1.1723&  1.1713\\
12&	1.0853&	1.1661&  1.1660\\
16&	1.0809&	1.1640&  1.1670\\
20&	1.0774&	1.1662&  1.1693\\
24&	1.0755&	1.1692&  1.1718\\
28&	1.0743&	1.1735&  1.1751\\
32&	1.0721&	1.1764&  1.1763\\
36&	1.0719&	1.1820&  1.1804\\
40&	1.0711&	1.1855&  1.1832\\
\hline
\end{tabular}
\caption{The numerically computed values of $E/B$ for 
$B=1$ and $B=4n$ with $1\le n\le 10$. 
The first column is the baryon number. The second column is that computed 
in a full field theory simulation with $m=0$, while the third column
 is the equivalent for $m=m_*$. The final column is that computed for 
$m=m_*$ using the rational map prediction (\ref{prediction}) and the data
in the second column. Notice that the third and fourth columns are very 
close to each other, the differences typically being in the third 
decimal place.}
\label{tab-4n}
\end{table}

\begin{figure}[ht]
\begin{center}
\leavevmode
\vskip -3cm
\epsfxsize=15cm\epsffile{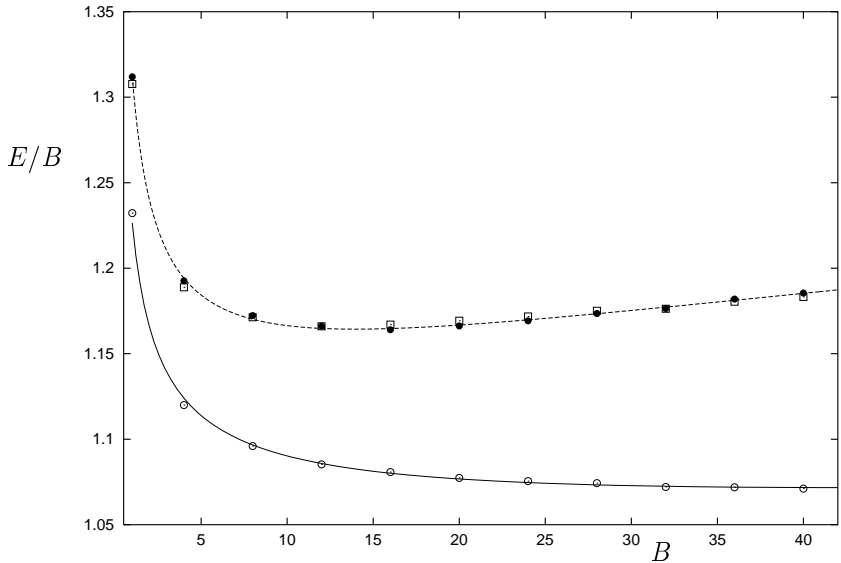}
\vskip -10.5cm
\caption{The values of $E/B$ presented in Table 1 plotted against $B$. The open
 circles are the second column, the solid squares are the third column and the 
open squares are the final column. 
Included also are the two fits discussed in section 3.}
\label{fig-4n}
\vskip 0cm
\end{center}
\end{figure}

From the point of view of our discussion here, the most important feature
 of the
 shell-like structure of Skyrmions is that the value of the field is close 
to $U=-1$
 in the centre. Since $U=1$ at spatial infinity the Skyrmion can be thought
 of as 
a novel spherical domain wall solution, with only a finite point group
 symmetry, 
interpolating between $1$ and $-1$. This is important since the term in
 the energy
 density associated with the pion mass is non-zero inside the shell and hence
 the 
overall energy will contain an extra term which will scale with the volume
 enclosed
 by the shell. If the windings associated with the topological charge are 
associated
 only with the shell, then the baryon number $B$ will scale like the surface
 area
 of the shell, and hence it is easy to see that there will be a term in
 $E/B$ which
 is proportional to $\sqrt{B},$ and therefore it grows with $B$.

This is very much in contrast to the case of $m=0$ for which all evidence
 appears
 to suggest that the value of $E/B$ tends to some asymptotic value 
which is only a few percent above the Faddeev-Bognomolny bound of $E/B \ge 1$. 
If the above discussion is correct then there is one important consequence: 
for sufficiently large values of $B$ the shell-like solutions will
 be unbound to a decay into smaller shells for any non-zero value of $m$. 
Hence, we must conclude that there are either another class of solutions 
for which $E/B$ tends to a constant for large $B$, or more radically that 
there are no bound static solutions for very large $B$.

In order to test this radical proposal we have used the same methods as 
developed 
in Ref.\cite{BS3} to compute accurate numerical energies of shell-like
 solutions 
with $m\ne 0$, concentrating initially on the case of $m=m_*$ for $B=4n$ with
 $1\le n\le 10$. We use an initial configuration created using the 
rational map ansatz
 with the symmetry thought to be that of minimal energy~\cite{BS3}. 
This was then 
relaxed toward a minimum using numerical field theory methods on a parallel
 supercomputer in 3D to find the numerical solution. We note that 
imposition of the exact symmetry from the beginning will ensure that, 
even if the
 solution is unstable, it will last sufficiently long within our simulations
 for an accurate energy to be computed.

The results of these simulations are presented in Table 1 and
 Fig.~\ref{fig-4n}.
 We see that for $m=0$ and $B\le 20$ the computed values of $E/B$ are in good
 agreement with the previously presented values \cite{BS3} 
and that values for large $B$
 are consistent with the asymptote of $\approx 1.061$
suggested by the Skyrme lattice \cite{BS4}.
 There is clearly a very
 different asymptotic behaviour when $m=m_*$ with $E/B$ appearing to increase 
with $B,$ for $B>16.$

For the range of baryon numbers we are considering, 
the energies per baryon presented in Fig.~\ref{fig-4n}
can be well approximated by a fit of the form
\be
\frac{E}{B}=\frac{\alpha}{\sqrt{B}}+\beta+\gamma\sqrt{B}.
\label{fit}
\ee
For $m=0$ the best fitting 
parameters are $\alpha=0.215,\ \beta=1.007, \gamma=0.005$
and for $m=m_*$ they are $\alpha=0.276,\ \beta=1.016, \gamma=0.020.$
These fits are shown as the curves in  Fig.~\ref{fig-4n}.
The most substantial difference between these two sets
of parameters is the fourfold increase in $\gamma$ with the inclusion
of the pion mass. As discussed above 
this term has the interpretation of a volume 
contribution to the energy, and this increase appears to uphold the
 associated explanation. It is important
to note that the fit (\ref{fit}) is only appropriate for the range
of values currently under consideration, that is, $B\le 40.$
It is not necessarily appropriate for larger values of $B,$ where, for
example, in the case of massless pions we expect an asymptotic
approach to the constant value $1.061$ associated with large shells. 

For values of $B$ beyond the minimum in $E/B$ for $m=m_*$ the possibility
 of the corresponding shell-like solution being unbound arises. On inspection
 of the computed energies we see that the energy of the $B=24$ shell
 solution is greater than that of two with $B=12.$ Hence we conclude that 
(at least) for $B=24,$ and probably all $B$ beyond this, the shell 
solutions are unbound for this value of $m$. 
It is our expectation that the critical value of $B$ at which
 the solutions become unbound is smaller for larger values of $m$.

The pion mass term considered in this paper is linear in the Skyrme field and
this is the one that is generally used in the literature. However, this term
is only really determined upto quadratic order in the pion fields so
alternative possibilities exist, such as a pion mass term which is
quadratic in the Skyrme field. Our results suggest that there will be
significant qualitative differences between these two types of pion mass term,
since a term quadratic in the Skyrme field does not penalize the region
where $U=-1$ so shells should again be favoured as in the massless pion
theory.

\section{Thirytwo Skyrmions: Shell vs Crystal}\news\label{sec-32}

\begin{figure}[ht]
\begin{center}
\leavevmode
\vskip -0cm
\epsfxsize=14cm\epsffile{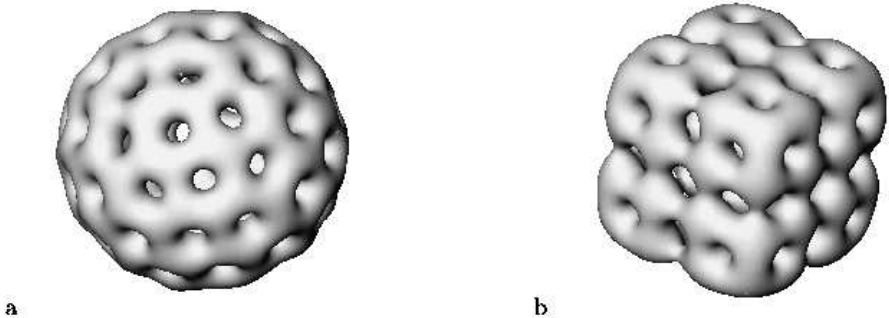}
\vskip -0cm
\caption{Baryon density isosurfaces for charge $32$ Skyrmions with massless
pions; (a) the $D_5$ shell, (b) the cubic crystal chunk.}
\label{fig-32sandc}
\vskip 0cm
\end{center}
\end{figure}

In the previous section we have presented some numerical evidence that 
shell-like solutions do not have minimal energy if $m=m_*$ for $B\ge 24$. 
One possibility is that the solutions become like chunks of the Skyrme crystal
 discussed earlier. Natural chunks of the crystal, with cubic symmetry,
 can be constructed with $B=4n^3$ for $n=1,2,..$ using the method discussed
 below. In this section we compute $E/B$ as a function of $m$ for a 
solution which is constructed from 
the minimal energy rational map with degree 32 and for a chunk of 
the Skyrme crystal with $B=32$. It is clear that the value of $E/B$ for a 
chunk of the Skyrme crystal will not have the same properties as
 shell-like solutions since the baryon number will scale like volume 
in constrast to the surface area dependence in the case of shells.

For $B=32$ the minimal energy rational map has $\I=1297.3=1.2668\times 32^2.$
It has the dihedral symmetry $D_5$ and is given by
\be
R=\frac{a_1z^2+a_2z^7+a_3z^{12}+a_4z^{17}+a_5z^{22}+a_6z^{27}+z^{32}}
{1+a_6z^5+a_5z^{10}+a_4z^{15}+a_3z^{20}+a_2z^{25}+a_1z^{30}},
\ee
where $a_1=-0.1-i33.3,\
a_2=1992.1+i906.8,\ 
a_3=312.0+i20441.1,\
a_4=24183.5+i8826.7,\
a_5=-8633.2-i4867.6,\
a_6=383.3-i559.6.$

To create a Skyrme field with baryon number 32 from a chunk of the 
Skyrme crystal we follow the approach of Ref.\cite{Ba}, in which
an interior portion of the crystal is cut out and the fields at the
boundary of this chunk are interpolated to the vacuum configuration.

The approximate formulae (\ref{crystal}) are used to describe
the fields of the Skyrme crystal, though it is first convenient to perform
a chiral $SO(4)$ rotation and work with the rotated fields
\be
\widetilde\sigma =\frac{1}{\sqrt{3}}(\pi_1+\pi_2+\pi_3),\quad
\widetilde\pi_i=-\frac{\sigma}{\sqrt{3}}-\frac{1}{3}(\pi_1+\pi_2+\pi_3)
+\pi_i.
\ee
A cutoff $\lambda$ is chosen, which is slightly less than the crystal
cell period $2L,$ and within the cube $-\lambda\le x_i\le \lambda$ the
above rotated fields are used. At a point ${\bf x}$ outside this cube, 
the direction of the pion fields is fixed to the value taken at the point 
where the half-line from the origin through ${\bf x}$ meets the surface of the
 cube. The $\sigma$ field outside the cube is determined by a simple
linear profile function of the distance from the cube's surface, which
interpolates between the value on the surface and the vacuum
value $\sigma=1$ at the boundary of the grid.

\begin{figure}[ht]
\begin{center}
\leavevmode
\vskip -1cm
\ \hskip -5cm
\epsfxsize=18cm\epsffile{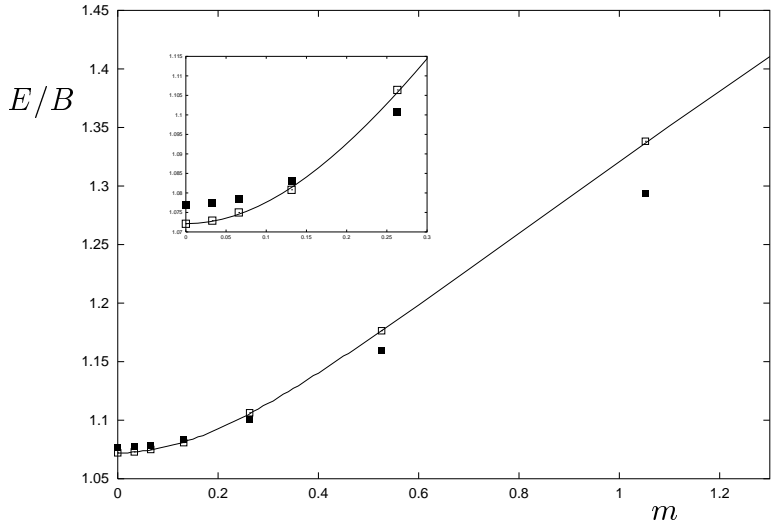}
\vskip -18cm
\caption{The energy per baryon for the charge 32 shell (open squares)
and the charge 32 crystal chunk (solid squares) as a function of the pion
mass. The curve is the rational map predicted energy for the 
charge 32 shell. The inset is a magnification of the first part
of the graph.}
\label{fig-32}
\vskip 0cm
\end{center}
\end{figure}

The results of our numerical relaxations are presented in Fig.~\ref{fig-32}
 for $m=0$ and $m=2^{n-5}m_*$ for $1\le n \le 6$. We see that for low values
 of $m$ the shell-like solution created from the rational map is the 
lowest energy solution, but for larger values of $m$ the crystal chunk is 
clearly much more tightly bound. The cross-over from a shell to the 
chunk of crystal being lower energy  appears to take place around $m\approx 0.16$. We should note that this does not by itself prove that the minimum energy Skyrmion is a chunk of the crystal, but we have shown that there is a solution with lower energy than the best shell.

It is interesting to note that for most of the range of $m$ displayed in
Fig.~\ref{fig-32} the behaviour of $E/B$ indicates that $E/B\propto m,$
and only for very small $m$ is the quadratic $E/B\propto m^2$ 
behaviour seen, which is expected from the naive examination of the 
contributions to the energy density. 
It is clear that there is an important energy dependence on $m$ which 
comes from a modification to the profile of the Skyrmion. 
In fact, within the rational map 
description, it is possible to show that for very large $m$ then
the profile function is deformed so that
$E/B\propto \sqrt{m},$ and this is evident in the figures displayed in
the following section, where we study large pion masses for low baryon numbers.

\begin{figure}[ht]
\begin{center}
\leavevmode
\vskip -1cm
\ \hskip -0cm
\epsfxsize=15cm\epsffile{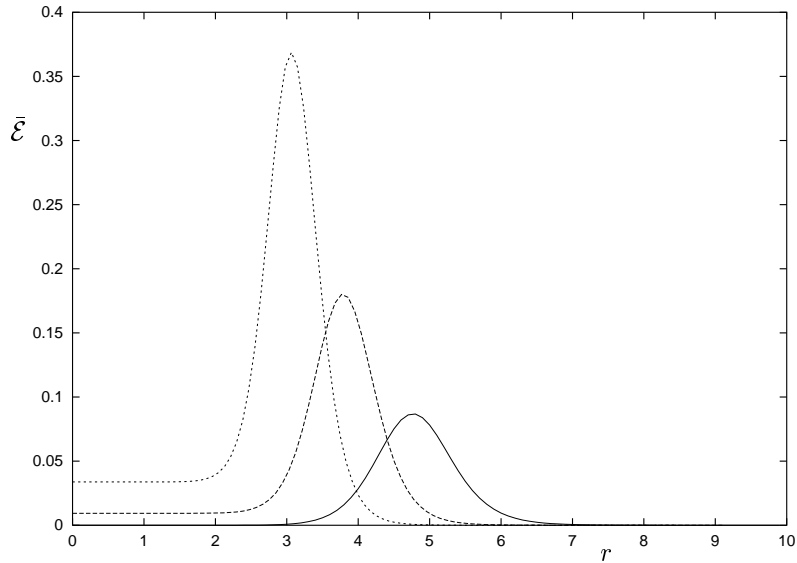}
\vskip -13.5cm
\caption{The energy density averaged over the sphere of radius $r$
for the charge 32 shell, using the rational map ansatz; $m=0$ (solid curve),
$m=m_*$ (dashed curve), $m=1$ (dotted curve).}
\label{fig-den32}
\vskip 0cm
\end{center}
\end{figure}
As we discussed in the previous section, the term in the energy density
associated with the pion mass will produce a total contribution to the
energy of a shell-like Skyrmion which scales like the volume enclosed 
by the shell. In Fig.~\ref{fig-den32} we display this effect by
plotting the energy density averaged over the sphere of radius $r$
for the charge 32 shell. These results were obtained using the rational
map ansatz. The solid curve is the massless pion case $m=0,$ where the
energy density is localized around the shell. The dashed curve is
for $m=m_*$ and it can be seen that although the energy density is
still peaked around a shell it attains a constant value inside the shell,
producing a volume contribution to the energy. As the pion mass is increased
(the dotted curve corresponds to $m=1$) the density inside
the shell increases and the radius of the shell decreases to compensate.

\section{Large pion masses}\news\label{sec-large}
As discussed in the introduction, the value of the pion mass parameter
in the Skyrme model depends not only on the experimentally measured pion mass,
but also on how the other parameters of the Skyrme model are fixed by
comparison with experimental data. In recent years there has been a substantial
increase in the results available for multi-Skyrmions and so it is perhaps
time to reconsider alternative approaches to fitting the Skyrme parameters
to experimental data, with a view to modelling nuclei of all baryon numbers,
rather than restricting attention only to the single baryon to fit
the Skyrme parameters, as is done in Refs.\cite{ANW,AN} from which
the standard value originates. This motivates our investigations in this
section, where we consider shell-like Skyrmions with baryon
numbers one to eight, and pion masses from zero upto a value several times
larger than the standard one.
\begin{figure}[ht]
\begin{center}
\leavevmode
\vskip -2cm
\epsfxsize=12cm\epsffile{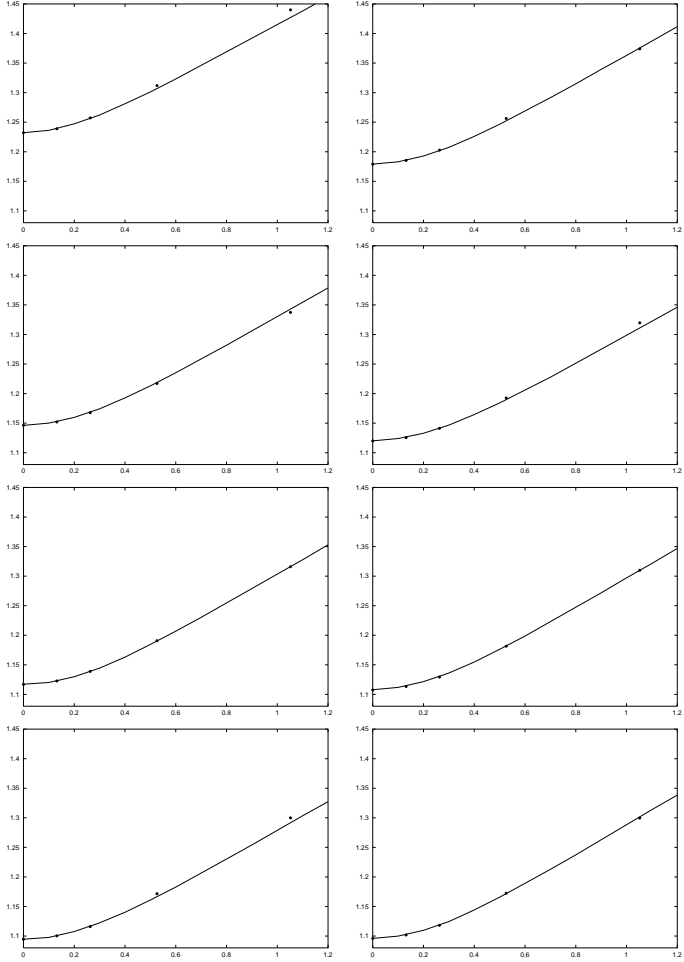}
\vskip -3.5cm
\caption{The energy per baryon for $1\le B\le 8,$ as a function of the 
pion mass in the range $0\le m\le 1.2.$
 Baryon number increases for the figures from left to
right and then top to bottom. Solid circles are the results of
full field simulations and curves are the rational map predictions.}
\label{fig-stan_all8}
\vskip 0cm
\end{center}
\end{figure}

\begin{figure}[ht]
\begin{center}
\leavevmode
\vskip -1cm
\epsfxsize=15cm\epsffile{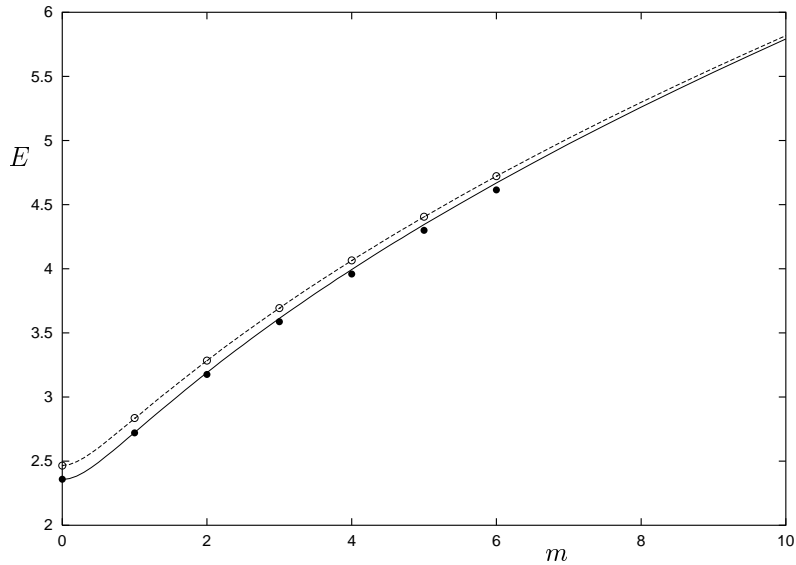}
\vskip -13.5cm
\caption{The energy of the $B=2$ Skyrmion (solid curve and solid circles)
and twice the energy of the $B=1$ Skyrmion (dashed curve and open circles).
The circles are the full field simulations assuming axial symmetry 
and curves are the rational map
predictions.}
\label{fig-bind2}
\vskip 0cm
\end{center}
\end{figure}

\begin{figure}[ht]
\begin{center}
\leavevmode
\vskip -1cm
\epsfxsize=15cm\epsffile{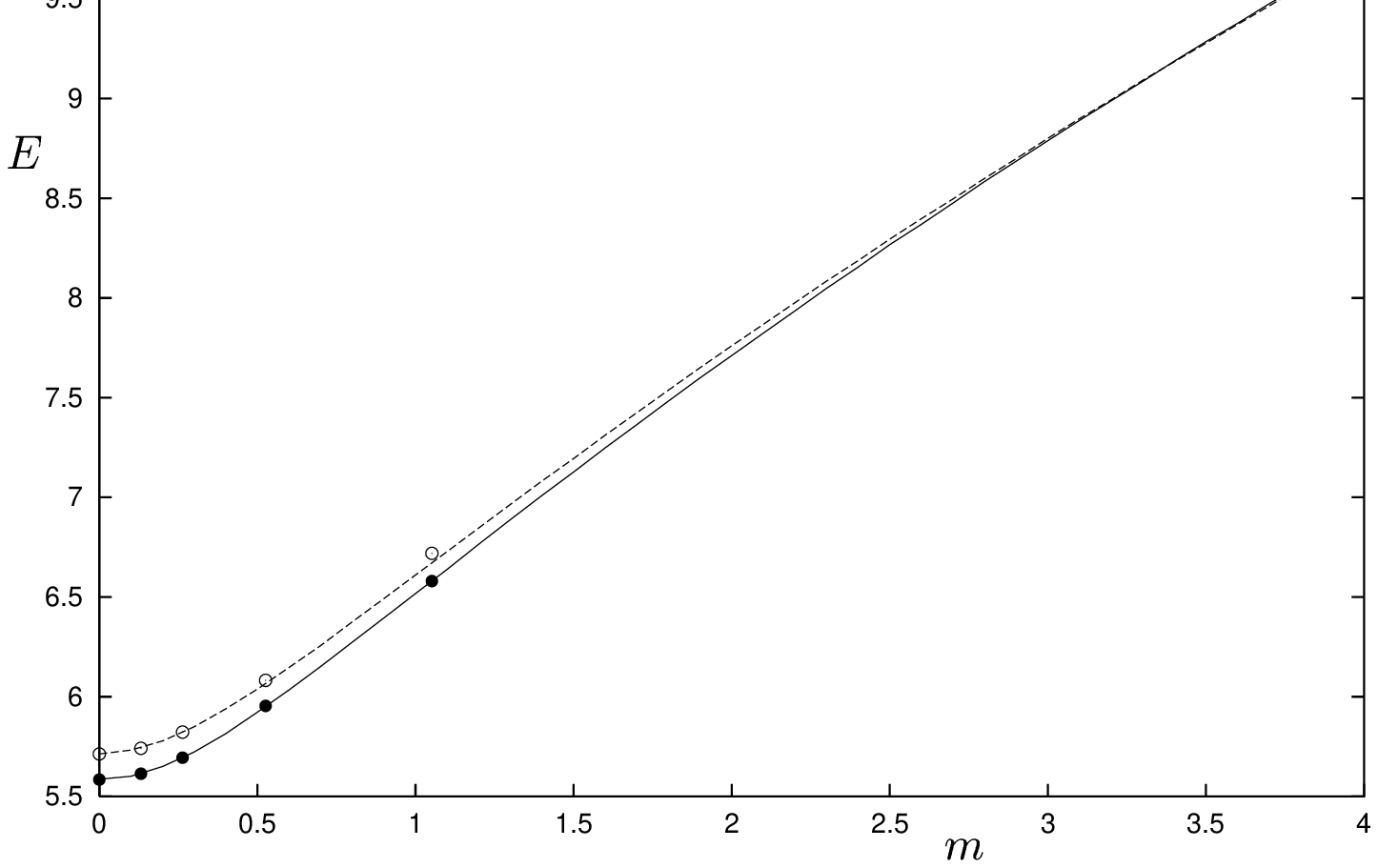}
\vskip -13.5cm
\caption{The energy of the $B=5$ Skyrmion (solid curve and solid circles)
and the energy of the $B=4$ plus $B=1$ Skyrmions 
(dashed curve and open circles).
The circles are the full field simulations and curves are the rational map
predictions. }
\label{fig-bind5}
\vskip 0cm
\end{center}
\end{figure}

\begin{figure}[ht]
\begin{center}
\leavevmode
\vskip -1cm
\epsfxsize=15cm\epsffile{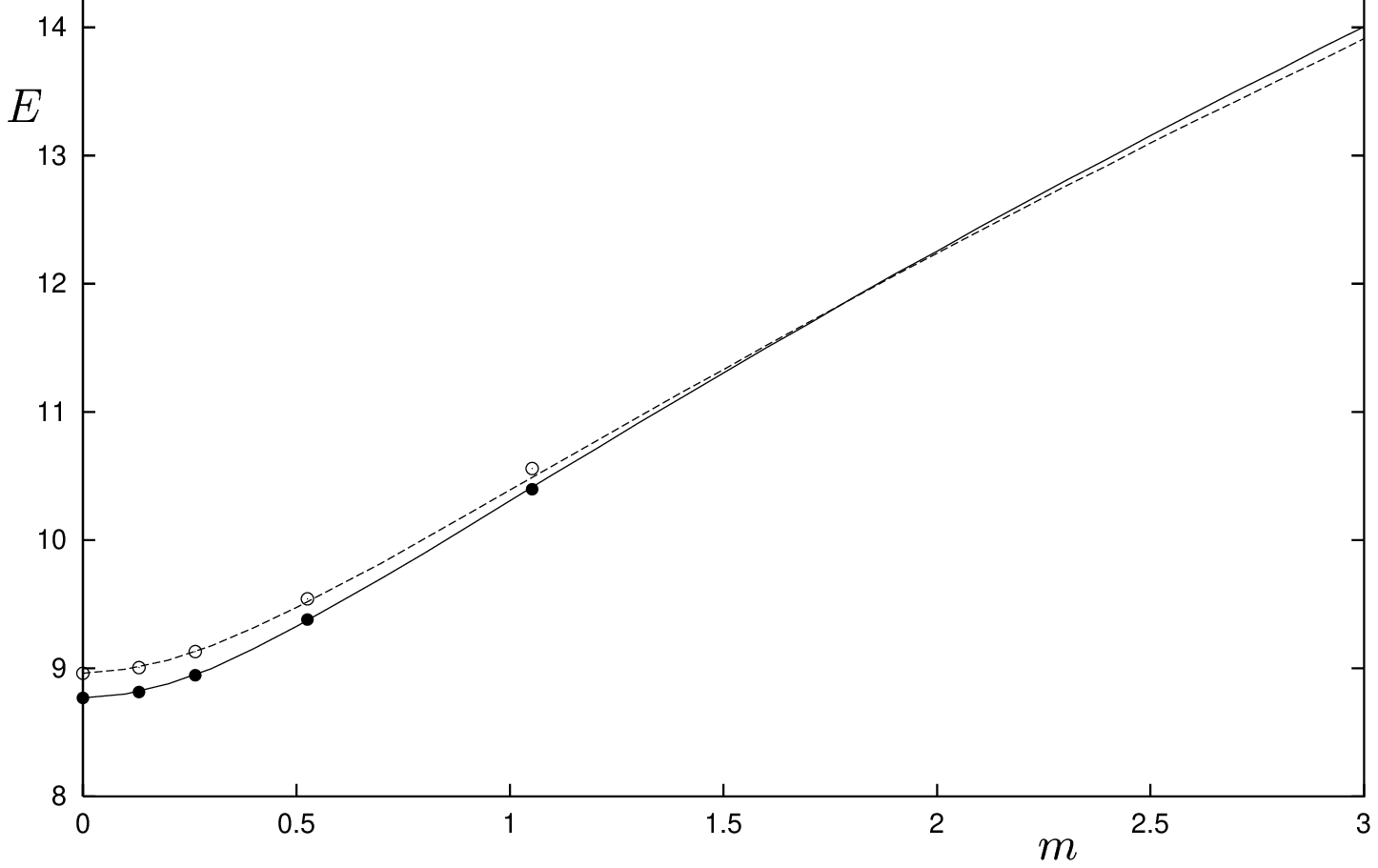}
\vskip -13.5cm
\caption{The energy of the $B=8$ Skyrmion (solid curve and solid circles)
and twice the energy of the $B=4$ Skyrmion (dashed curve and open circles).
The circles are the full field simulations and curves are the rational map
preidctions. }
\label{fig-bind8}
\vskip 0cm
\end{center}
\end{figure}

In Fig.~\ref{fig-stan_all8} we plot (solid circles) 
the results of full field simulations to compute the energy per baryon
with $1\le B\le 8$ for a range of pion masses from zero upto twice the 
standard value. The curves are the rational map predicted energies using
the formula (\ref{prediction}). As can be seen from this figure, there
is a good agreement between these two different methods, which provides
evidence for the validity of both. 

Recall that for $B=1$ the rational map ansatz produces the exact
solution, so in this case the curve gives the true value. The profile
function must be computed numerically, but since this requires the
solution of only an ordinary differential equation this can be done
with a high accuracy. It is curious to note that the largest discrepancy
(though it is still quite small) between 
the rational map prediction and the full field simulations occurs
for $B=1,$ with the pion mass being the largest value considered.
As the rational map approximation is exact in this case then we know
that the error here is in the full field simulation. The reason for this 
error is that it becomes more difficult to accurately calculate the energy in a
full field computation as the pion mass increases because the fields 
have a more rapid spatial variation. This effect is greatest for $B=1$
because this is the smallest Skyrmion, so it is the most localized.
The increased localization with larger pion mass means that we are unable
to determine accurate energy values much beyond twice the standard pion
mass using full field simulations. However, since the rational map
predicted energies are in such good agreement with the full field computations
this gives us confidence to trust the rational map predicted energies for
larger pion masses.

As a further test of the rational map energies, we can compare with 
exact calculations of the energy 
of the $B=2$ Skyrmion. The $B=2$ Skyrmion is axially
symmetric, so the computation is effectively 2D and this allows larger
grids to be used than in 3D, producing accurate results for larger pion masses.
The axial 2D computation was done using the method described in Ref.~\cite{KS}
and the results are shown as the solid circles in Fig.~\ref{fig-bind2}.
For comparison, the open circles show twice the energy of the $B=1$ Skyrmion,
 computed using the same axial code. The curves are the rational map
predicted energies for both these quantities. This figure supports the
use of the rational map energies even for very large pion masses, with the
errors still being small for pion masses upto ten times the standard value,
though beyond this even the accuracy of the 2D simulations 
is not good enough to make quantitative
statements.

Fig.~\ref{fig-bind2} suggests that the binding energy of the $B=2$
Skyrmion decreases with increasing pion mass. In fact the rational map
prediction is that for a large enough pion mass the $B=2$ Skyrmion is
not bound into the break-up to two single Skyrmions. However, the 
pion mass at which this occurs is so large that the rational map
energies can no longer be trusted, so it remains an open problem
as to whether the axial $B=2$ Skyrmion is bound for all pion masses.

For massless pions the $B=5$ and $B=8$ Skyrmions have relatively
low binding energies. This is an encouraging result of the Skyrme
model since there are no real nuclei which are stable with five
or eight nucleons. There is therefore a possibility that the
quantization of these classical solutions may reproduce the experimental
result of no bound states. However, let us investigate the classical results
further and consider how the binding energy of these shell Skyrmions
varies with the pion mass. 

In Fig.~\ref{fig-bind5} we display, as a function of the pion mass,
 the energy of the $B=5$ Skyrmion and
for comparison the energy of the $B=4$ Skyrmion plus the $B=1$ Skyrmion.
As in Fig.~\ref{fig-bind2}, and all the subsequent figures in this section,
the solid and open circles are the numerical field theory simulations
(computed upto pion masses at which the accuracy is reliable) and the
lines are the rational map predictions. In Fig.~\ref{fig-bind5} the
solid curve ($B=5$) is below the dashed curve ($B=4+1$) for low pion masses,
indicating that the $B=5$ Skyrmion is bound against losing a single Skyrmion,
but at $m\approx 3.3$ the curves cross, so that the $B=5$ Skyrmion
is no longer a bound state above this pion mass. 
Assuming the rational map energy is accurate upto these values (for which
we have presented reasonable evidence) this still does not prove that
there is no stable $B=5$ Skyrmion, but it does show that if a stable
Skyrmion exists then it can not be of the simple shell type.

It is interesting that the experimental result of no stable baryon number
five nucleus can be reproduced in the classical Skyrme model
if the pion mass is set to be much larger than the standard value.
This prompts us to examine the similar situation for $B=8.$
In Fig.~\ref{fig-bind8} we display
 the energy of the $B=8$ Skyrmion (solid curve) and
for comparison twice the energy of the $B=4$ Skyrmion (dashed curve).
The results are similar, with the $B=8$ Skyrmion becoming unbound at
an even smaller pion mass $m\approx 1.8$

We have checked all other possible decay combinations for the $B=5$ and
$B=8$ Skyrmions and found that the two described above are the 
relevant ones. Furthermore, we have examined all Skyrmions with $B\le 8$
for all pion masses $m\le 4$ and found that only the $B=5$ and $B=8$
Skyrmions are unbound in this range. Thus it appears a possibility
that there are values of the pion mass at which $B=5$ and $B=8$ 
are the only low charge Skyrmions for which there is no bound state.
For larger baryon numbers we expect shell Skyrmions to be replaced
by other configurations, as we have seen in detail for $B=32,$ but
for small values of $B$ it is not as clear that there are alternative
bound state configurations that appear at large pion mass. 
Whichever of these possibilities arises it is clearly an interesting
phenomenon which requires further investigation.

\section{Conclusion}\news\label{sec-conclusion}
We have studied some aspects of Skyrmions with a pion mass using both full 
field simulations and the rational map approximation, with the results 
from these two different approaches being in good agreement. These results
suggest that the introduction of a non-zero pion mass might dramatically
change both the quantitative and qualitative features of minimal energy
Skyrmions. In particular, shell-like Skyrmions which are minimal
energy bound states for massless pions fail to be bound states for 
massive pions, and we have presented some evidence that crystal chunks
may play an important role in the structure of minimal energy Skyrmions
with a pion mass.

The replacement of shell Skyrmions with crystal chunks, or some
similar more three-dimensional structure, 
has an important consequence which
improves a qualitative feature of the Skyrme model in comparison with
experimental data. Namely, for large baryon number $B$ the size of 
nuclei scales like $B^{1/3},$ which is observed experimentally, whereas
for shell-like Skyrmions the size grows like $\sqrt{B}.$ Thus the
inclusion of the pion mass is clearly important in comparing the 
results of the Skyrme model with the properties of real nuclei.

As the pion mass is increased its effects become important even for
very low baryon numbers and it appears that shell-like bound states
may fail to exist for baryon numbers five and eight. This is an interesting
feature since it reproduces the experimental result that there are no
stable nuclei with five or eight nucleons. 

In this paper we have demonstrated that interesting phenomena arise
when the pion mass is included in the Skyrme model. It is clearly
desirable to know more about the details of Skyrmions with massive
pions and further numerical investigations are currently in progress.
In a zero mode quantization of Skyrmions the symmetry of the classical 
solution plays an important role in providing constraints on the 
spin, isospin and parity quantum numbers. This has been studied in
detail \cite{Kr2} for $B\le 22$ with massless pions, with some limited
success in comparing with experimental data. However, 
the results we have presented suggest that the symmetries
of Skyrmions will be quite different for massive pions, so it is of
interest to determine these symmetries and calculate the new constraints
on quantum numbers that arise.

Finally, we have shown that a number of Skyrmion properties appear
to be quite sensitive to the value of the pion mass, so a detailed
study should provide the data necessary to reconsider the way in which
the parameters of the Skyrme model are set by experimental data. 
In particular, with a view to try and model nuclei of all baryon numbers
it would seem sensible to fit the parameters of the Skyrme model
by comparison with multi-Skyrmions over a range of baryon numbers,
rather than restricting the fit to a single Skyrmion.

\section*{Acknowledgements}
RAB acknowledges PPARC for an Advanced Fellowship.

\end{document}